\documentclass{article}

 \usepackage[preprint,nonatbib]{neurips_2025}

\usepackage[utf8]{inputenc} 
\usepackage[T1]{fontenc}    
\usepackage{url}            
\usepackage{booktabs}       
\usepackage{amsfonts}       
\usepackage{nicefrac}       
\usepackage{microtype}      
\usepackage{xcolor}         

\usepackage{amsmath,amsfonts,bm}

\def\eqref#1{equation~\ref{#1}}

\def\1{\bm{1}}

\DeclareMathAlphabet{\mathsfit}{\encodingdefault}{\sfdefault}{m}{sl}
\SetMathAlphabet{\mathsfit}{bold}{\encodingdefault}{\sfdefault}{bx}{n}

\usepackage{microtype}
\usepackage{graphicx}
\usepackage{subfig}
\usepackage{booktabs}   
\usepackage{amsmath,bm}
\usepackage{amsthm}
\usepackage{cite}
\usepackage{enumerate}
\usepackage{wrapfig}
\usepackage{color, soul}
\usepackage{psfrag}
\usepackage{adjustbox}
\usepackage{xspace}
\usepackage{amssymb}
\usepackage{multirow}
\usepackage{afterpage}
\usepackage{colortbl}
\usepackage{float}
\usepackage{textcomp}
\usepackage{siunitx}
\usepackage{mdframed}
\usepackage[super]{nth}

\usepackage{enumitem}
\usepackage{bbding}
\usepackage{pifont}
\usepackage{bbm}
\usepackage{xcolor}
\usepackage{tcolorbox}
\usepackage{theoremref}
\usepackage{pifont}
\usepackage{adjustbox}
\usepackage{pdflscape}
\usepackage{todonotes}
\usepackage[super]{nth}
\colorlet{darkgreen}{green!65!black}
\colorlet{darkblue}{blue!75!black}
\colorlet{darkred}{red!80!black}
\definecolor{statistical}{HTML}{8c564b}
\definecolor{structural}{HTML}{0070C0}
\definecolor{semantic}{HTML}{008080}
\definecolor{yellow}{HTML}{f7c600}
\definecolor{lightblue}{HTML}{0071bc}
\definecolor{lightgreen}{HTML}{39b54a}
\definecolor{deemph}{gray}{0.55}

\definecolor{baselinecolor}{gray}{.95}
\definecolor{graycolor}{gray}{.95}

\usepackage[pagebackref=false, breaklinks=true, colorlinks, citecolor=lightblue, bookmarks=false]{hyperref}
\usepackage[capitalize,noabbrev]{cleveref}
\usepackage{tikz}
\usetikzlibrary{patterns}
\newcommand{\cmark}{\cellcolor{green!20}\checkmark}

\newlength\savewidth
\newcolumntype{x}[1]{>{\centering\arraybackslash}p{#1pt}}
\newcolumntype{y}[1]{>{\raggedright\arraybackslash}p{#1pt}}
\newcolumntype{z}[1]{>{\raggedleft\arraybackslash}p{#1pt}}

\newcolumntype{R}[2]{%
    >{\adjustbox{angle=#1,lap=\width-(#2)}\bgroup}%
    l%
    <{\egroup}%
}
\newcommand*\rot{\multicolumn{1}{R{45}{1em}}}

\newcommand{\MandatoryCircle}[1][0.5]{%
    \tikz[baseline=(char.base)]\node[shape=circle,inner sep=2pt,draw=darkgreen,fill=darkgreen,line width=1pt,scale=#1] (char) {\phantom{T}};\hspace{-1pt}
}

\newcommand{\railLogo}[1][1]{
    \includegraphics[width=0.35cm]{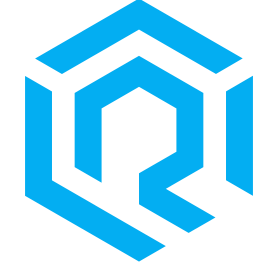}
}

\newcommand{\bigscienceLogo}[1][1]{
    \includegraphics[width=0.35cm]{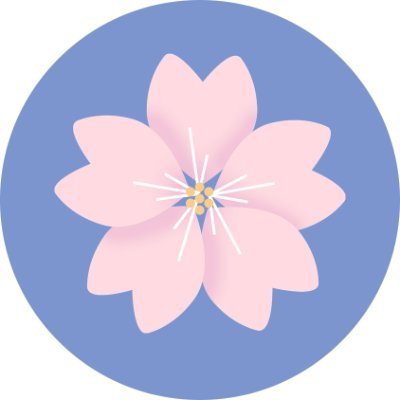}
}

\newcommand{\metaLogo}[1][1]{
    \includegraphics[width=0.35cm]{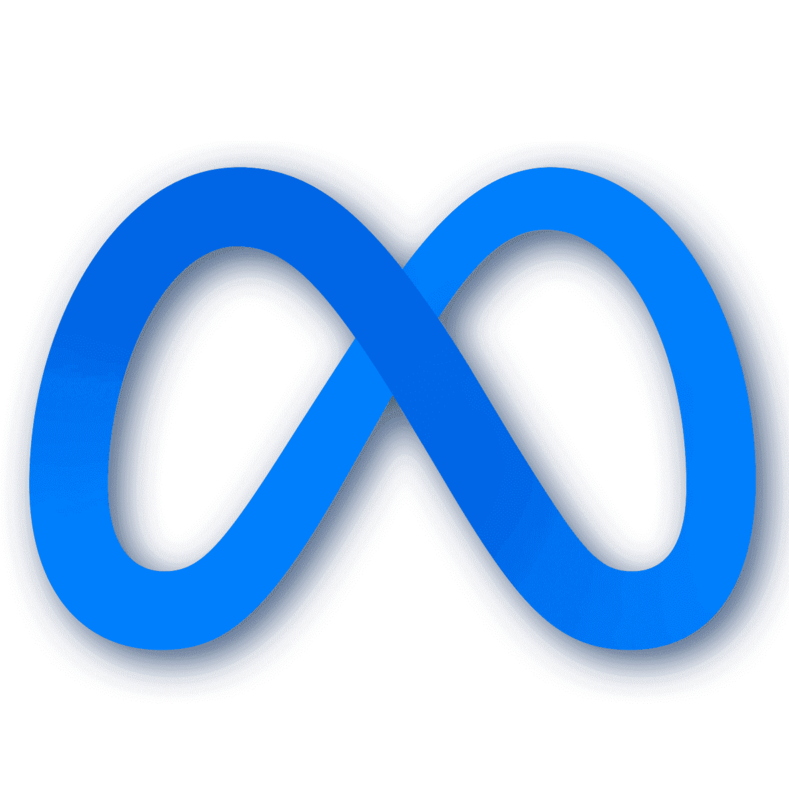}
}

\newcommand{\googleLogo}[1][1]{
    \includegraphics[width=0.35cm]{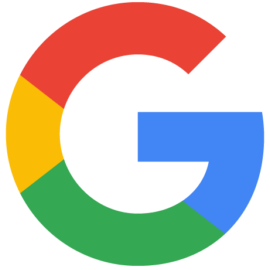}
}

\title{New Tools are Needed for Tracking Adherence to AI Model Behavioral Use Clauses}

\author{Daniel McDuff$^*$ \And Tim Korjakow \And Kevin Klyman \And Danish Contractor}

\begin{document}

\maketitle
\def\thefootnote{*}\footnotetext{Correspondence: \tt{dmcduff@uw.edu}}

\begin{abstract}
Foundation models have had a transformative impact on AI. A combination of large investments in research and development, growing sources of digital data for training, and architectures that scale with data and compute has led to models with powerful capabilities. Releasing assets is fundamental to scientific advancement and commercial enterprise. However, concerns over negligent or malicious uses of AI have led to the design of mechanisms to limit the risks of the technology. The result has been a proliferation of licenses with behavioral-use clauses and acceptable-use-policies that are increasingly being adopted by commonly used families of models (Llama~\cite{touvron2023llama}, Gemma~\cite{team2024gemma2}, Deepseek~\cite{liu2024deepseek_v3}) and a myriad of smaller projects. We created and deployed a custom AI licenses generator to facilitate license creation and have quantitatively and qualitatively analyzed over 300 customized licenses created with this tool. Alongside this we analyzed 1.7 million models licenses on the HuggingFace model hub. Our results show increasing adoption of these licenses, interest in tools that support their creation and a convergence on common clause configurations. \textbf{We take the position that tools for tracking adoption of, and adherence to, these licenses is the natural next step and urgently needed in order to ensure they have the desired impact of ensuring responsible use.}
\end{abstract}

\section{Introduction}

Rapid development and deployment of artificial intelligence (AI) have raised concerns about its potential misuses, such as the generation of harmful content, violating laws, and perpetuating biases. To address these concerns, researchers and developers have begun exploring ways to manage the risks associated with AI, including the use of licenses with behavioral use clauses \cite{contractor2022behavioral}. These licenses, sometimes referred to as Responsible AI Licenses (RAIL), allow developers to release AI assets while specifying restrictions on their use, thus promoting transparency and accountability and are a key tool in the responsible development of foundation models~\cite{longpreresponsible}. 

The adoption of RAIL licenses (or licenses with behaviorial-use clauses)\footnote{We make no distinction between licenses using Behavioral-use clauses and licenses using the RAIL acronym for this work.} has gained momentum in recent years, with notable examples including the models from the BigScience~\cite{workshop2022bloom} and BigCode~\cite{li2023starcoder} initiatives,  Stable Diffusion \cite{rombach2022high}, the Llama family of language models \cite{touvron2023llama}, Gemma family of language models \cite{team2024gemma,team2024gemma2}, as well as many of the models from DeepSeek~\cite{liu2024deepseek_v2,liu2024deepseek_v3}.  These licenses restrict the use of AI assets in various ways, such as prohibiting the generation of harmful content or the use of AI for malicious purposes~\cite{mcduff2024position}. 
As RAIL licenses have become more common regulatory bodies and think tanks have begun to encourage their use. The French Government's PEReN center of expertise, which serves digital regulation, expresses an explicit preference for models with a high degree of openness, but with a ``license which does not allow unethical usage''. ~\footnote{\url{https://www.peren.gouv.fr/en/compare-os-iag/}} 

Despite the growing adoption of Responsible AI Licenses, there is a need for further research on their effectiveness and limitations. Previous studies have highlighted the challenges of regulating AI, including the complexity of AI supply chains and the uncertainty surrounding copyright rules. Furthermore, \cite{mcduff2024position} argue that the lack of standardization in licenses with behavioral use clauses can lead to confusion and inconsistency and suggest the use of tooling to help with standardization. \cite{mcduff2024position}'s detailed study highlighted how the share of such licenses applied on models  grew from close to 0 to over 20\% in the 18-months from mid-2022. Adoption of such licenses was driven in part by major foundation model developers~\cite{touvron2023llama, team2024gemma,workshop2022bloom}.

The continued rise in the popularity of these licenses motivates our position in this paper. We take the position that \textbf{new tools are needed for tracking the adoption of, and adherence to, behavioral use clauses in AI licenses.} In this paper, we present two sets of analyses that provide evidence to motivate this position. The first is an in-depth examination of the usage of an open-source RAIL-License Generator proposed by McDuff et al.~\cite{mcduff2024position}. The license generator allows practitioners to select usage restrictions from a library of pre-populated clauses which are appended on a set of $10$ mandatory clauses identified by an analysis of existing RAIL licenses \cite{mcduff2024position}. Additionally, it allows specifying the artefacts being licensed (model, source code, or application) as well as the nature of release - research-use, open or proprietary (See Section \ref{sec:license_generator}). 
The second study is a large-scale analysis of licenses used by 1,704,180 models available on the HuggingFace model hub that highlights the continued large-scale adoption of RAIL licenses and the coherence of license clauses. 

Our studies provides a unique window into the practical application of RAIL licenses, revealing growing adoption and several canonical modes of adoption. By analyzing the selection and combination of behavioral-use clauses, we can identify areas of convergence and divergence in the community's understanding of responsible AI development and use. We study the licenses generated by users to gain a deeper understanding of how practitioners are selecting and applying these clauses. Our analysis reveals interesting trends and patterns in the selection of behavioral-use clauses, shedding light on the types of restrictions that users deem most important for responsible AI development and use. For example, we find that clauses related to disinformation/fake news were among the most frequently selected, indicating that people are particularly concerned that text and multimodal content generation models could be particularly dangerous for these purposes. In contrast, clauses related to warfare and injury or death  were less commonly chosen. We believe that these outcomes are not viewed as less damaging but rather much less likely outcomes from a generative foundation model. As embodied AI and robotics applications grow these may be represented more.

To summarize, while there are now tools to support the creation and customization of licenses, the evidence now supports the case for new tools to support practitioners to track \emph{how licenses are being adopted} and \emph{how license assets are being used}. Without such tools cynicism about the effectiveness and enforcability of such licenses might grow and violation of licenses, and the associated negative impacts, might increase. 
We argue that the availability, or unavailability, of such tools will have significant implications on the positive impact of Responsible AI Licenses. Furthermore, our findings can inform policymakers and regulators seeking to establish frameworks for the governance of AI, ensuring that these frameworks are grounded in the realities of AI development and use. 

\section{Case Study I: A RAIL License Generator}
\label{sec:license_generator}

\begin{figure}[t]
    \centering
    \includegraphics[width=\textwidth]{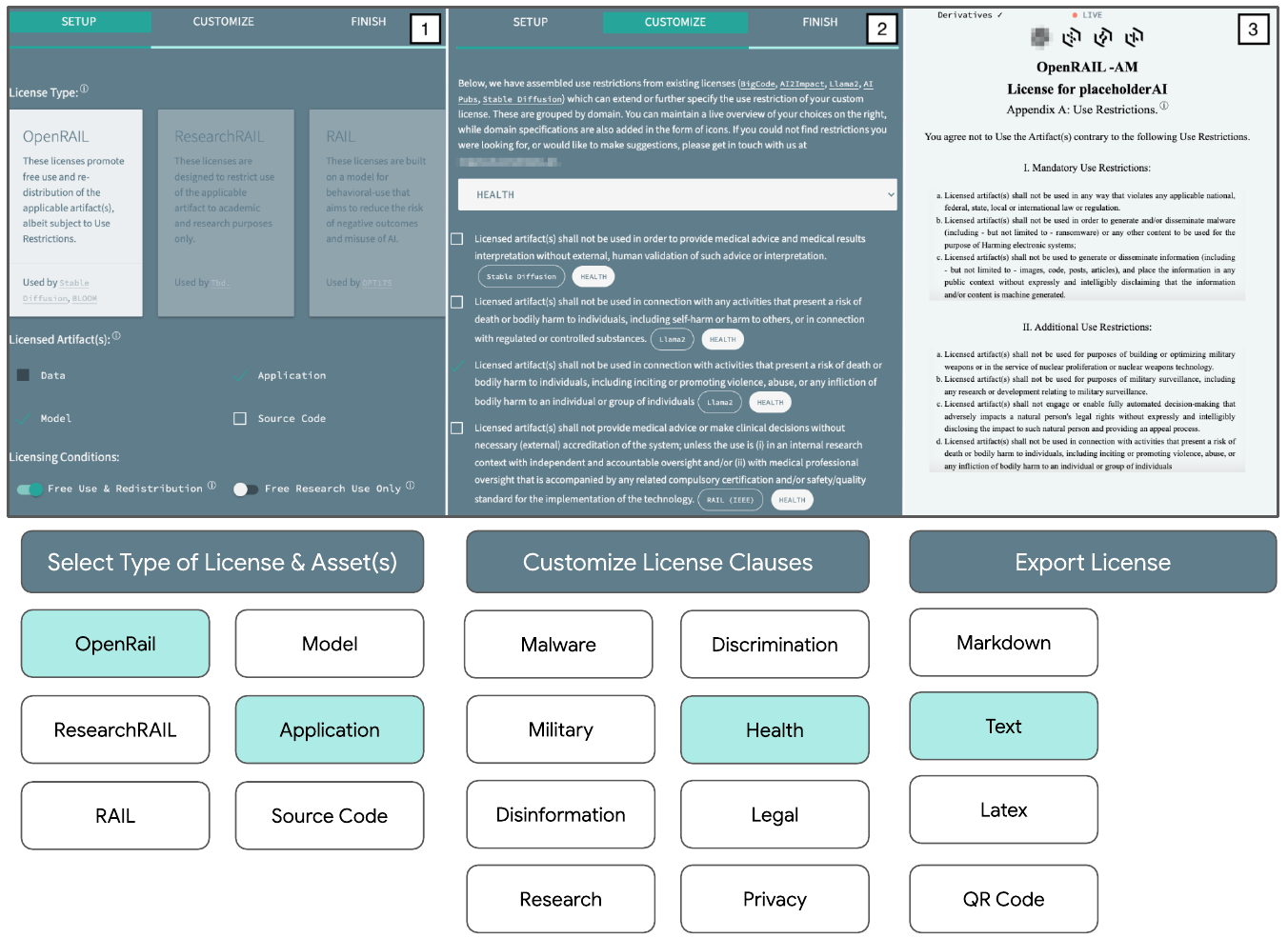} 
    \caption{\textbf{License Generator.} The RAIL License Generator enables a user to select a type of license and the asset they want to license and then customize the behavioral use clauses. The final license text can be exported in several ways along with a quick response (QR) code.}
    \label{fig:license_generator}
\end{figure}

We adopt the Open Source license generator proposed by McDuff et al.~\cite{mcduff2024position} for our work. The generator was hosted publicly at:~\url{https://www.licenses.ai/rail-license-generator}. The generator has a three-step workflow for generating licenses (see Fig.~\ref{fig:license_generator}). First, users are expected to indicate whether the license is being generated is for artifacts that can be used for research-only (ResearchRAIL), proprietary-use (RAIL) or open-use (OpenRAIL). Next, users select the artifact being licensed - model (M), source (S) code (C) or an application (A).  Existing behavioural-use clauses largely apply to models, though there are instances of some being applied to source code and applications. Finally, users are presented with a set of $10$ mandatory clauses and $15$ optional clauses that they can choose to include in their license (See Table \ref{tab:license_clauses}). We use the clauses identified previously by \cite{mcduff2024position} but as we report in Section \ref{sec:lg_analysis}, the adoption of clauses by the community suggests some revision may be necessary. 
The process of generating a license is summarized in Figure \ref{fig:license_generator}.

\begin{figure}[t]
\centering
\includegraphics[width=\textwidth]{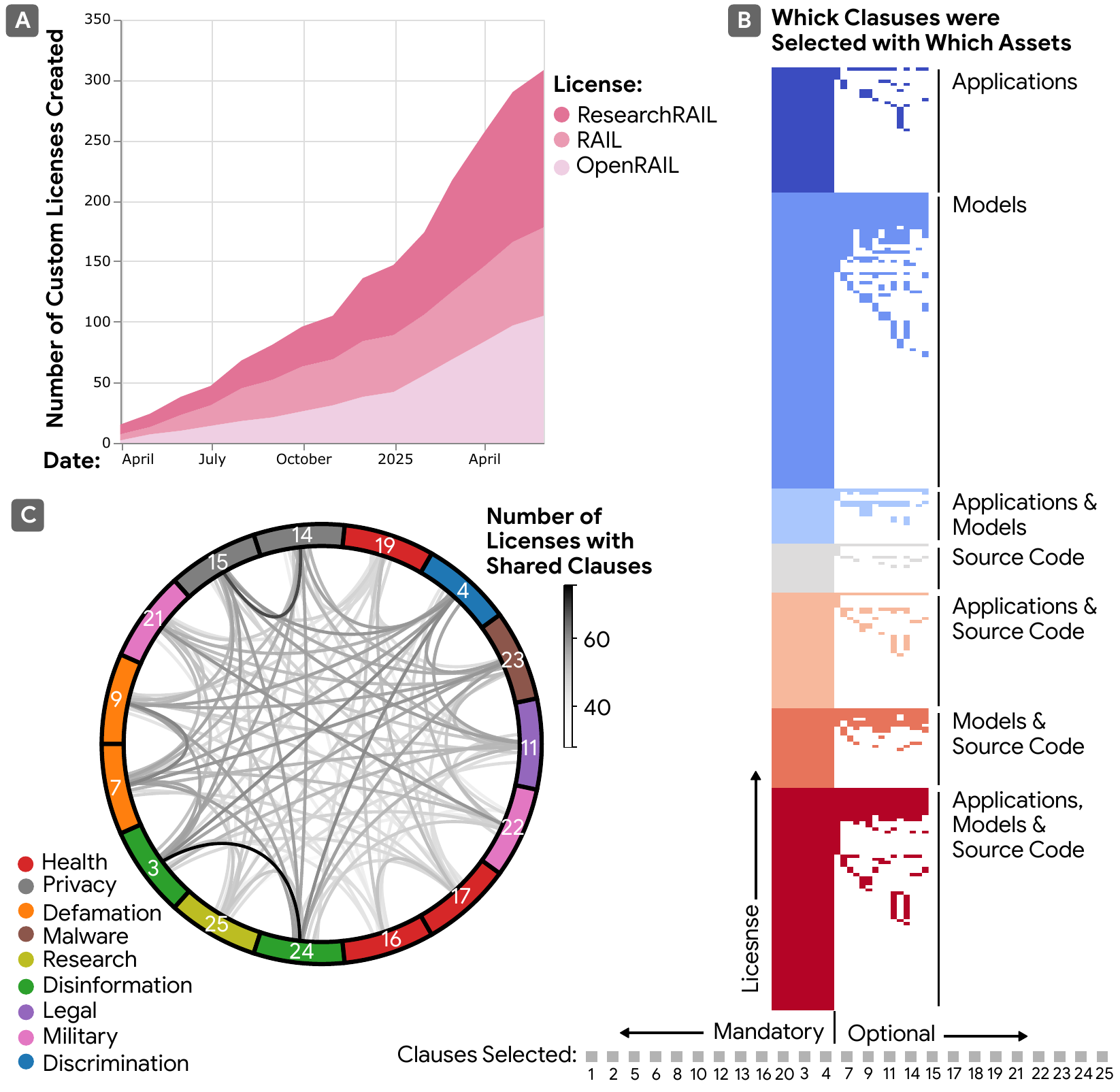}
\caption{\textbf{License Generator Usage.} 
[A] \textbf{Number of Licenses Created.} The number of RAIL, OpenRAIL and ResearchRAIL licenses created using our license generator has been accelerating over the year from April 2024. 
[B] \textbf{Clause Adoption by Asset Type.} Clauses select for every license generated for each type of asset. 
[C] \textbf{Connectivity Plot of Non-Mandatory Clauses.} The a circular plot highlights the heterogeneity across licenses in terms of the non-mandatory clauses selected. The opacity of each line reflects the number of licenses that include both clauses.}
\label{fig:license_generator_usage}
\end{figure}

The license generator was launched on March \nth{13} 2024 and the data used in the analysis for this paper was pulled on May \nth{16} 2025.
Figure~\ref{fig:license_generator_usage}[A] shows the cumulative number of licenses created each month over the past year. ResearchRAIL, RAIL and OpenRAIL are different "flavors" of RAIL licenses as described in~\cite{mcduff2024position}. Over a period of $14$ months we had over $300$ complete licenses generated. We exclude licenses in the analysis that were clearly generated for testing purposes, specifically those whose names contain the substring "test" or have an empty name. This exclusion reduces noise in the collected dataset and allows for a clearer picture during the analysis of the connections between restrictions. 

Some examples of artifacts licensed using licenses from the generator include projects from various modalities and use cases such as vision models in maritime settings \cite{noauthor_orcasoundorca-eye-aye_2024}, \cite{noauthor_salmon-computer-visionsalmon-computer-vision_2025}, autonomous driving \cite{florent_bartoccioni_vavim_2025}, earth observation \cite{vandal_global_2024},  medical analysis \cite{joshi2024factorizephys},
large language models \cite{jiang_liu_instella_2025}, \cite{ximeng_sun_instella-vl-1b_2025}, \cite{isobe_amd-hummingbird_2025} and 3D foundation models \cite{foundationaiteam2025cuberobloxview3d}.

\subsection{Architecture and Deployment}

The license generator was implemented using a FastAPI backend paired with a frontend composed exclusively of HTML and CSS. Persistent storage and template version management are handled by a PostgreSQL database. Each generated license includes the complete license text, user-selected restrictions and artifacts, a unique license identifier, and the Git commit hash corresponding to the specific template version used. Given that the codebase is open source, this design provides transparent and publicly verifiable provenance for each generated license. Furthermore, the inclusion of unique identifiers and immutable license records enables distribution not only in text form, but also through QR codes. The license generator has been publicly deployed, making it openly accessible to the general community.

\subsection{Analysis}
\label{sec:lg_analysis}

\textbf{Number of Licenses Created.} 
Figure \ref{fig:license_generator_usage}[A] shows the rate at which licenses were created using our tool.  With little additional marketing or publicity over 300 customized licenses were created in just over one year. The number of licenses created each week has also accelerated.  All three license types were popular, with ResearchRAIL licenses (that restrict to research-only user only) being the most commonly selected.

\textbf{License Types and Artifacts Selected.}
Figure \ref{fig:license_generator_usage}[B] summarizes how the license clauses were selected for different artifacts. The first $10$ clauses are the mandatory clauses and are therefore present in all licenses. Models are the most common asset that licenses were generated for. Of the $308$ licenses we find that more than $50$\% of the licenses included models as the artifact being licensed. Additionally, we find that the users of our license generator do find value in licensing other artifacts such as source code and end applications with behavioral-use restrictions. Notably, nearly $50$\% of the licenses included restrictions on the use of source code. This perhaps indicates that AI developers are also concerned about the inappropriate use of powerful AI source code though it is unclear to what extent this reflects a wider need for such AI License types, as existing licenses with behavioral-use restrictions have traditionally only been applied on models or end-use.  

\textbf{Artifacts and Clauses Selection}
The choice of licenses clauses does vary by asset type (see Figure~\ref{fig:license_generator_usage}[B]. We find that, on average, users tend to select more behavioral-use restrictions when licensing \emph{models} and the fewest when licensing \emph{source code}. When licensing applications and source-code, users appear to select more clauses as compared to only applications. This perhaps suggests that applications, because they are designed for more specific purposes, rather than more generic models/code, are viewed as needing fewer behavioral-use restrictions.

\textbf{Behavioral-use clauses selected by users.} The license generator had a minimum set of clauses indicated in Table~\ref{tab:license_clauses} by a green dot.  The most popular clauses related to users that intentional deceive or mislead, violate laws or create/disseminate malware. The generation of deceptive or misleading content (e.g., fake news) is an easily imaginable malicious application of a text generation model. 

The least selected were related to uses that have a connection with activities that present a risk of death or bodily harm.  It is likely that these were less frequently selected because it is harder to imagine how a text generation model would lead to such an application.  

Figure~\ref{fig:license_generator_usage}[C] shows shows a connectivity (circular) plot that reflects which non-mandatory clauses were selected together.

\section{Case Study II: Commonly Adopted AI Licenses}\label{sec:commonlyadoptedailicenses}
\begin{figure}
\centering
\includegraphics[scale=0.5]{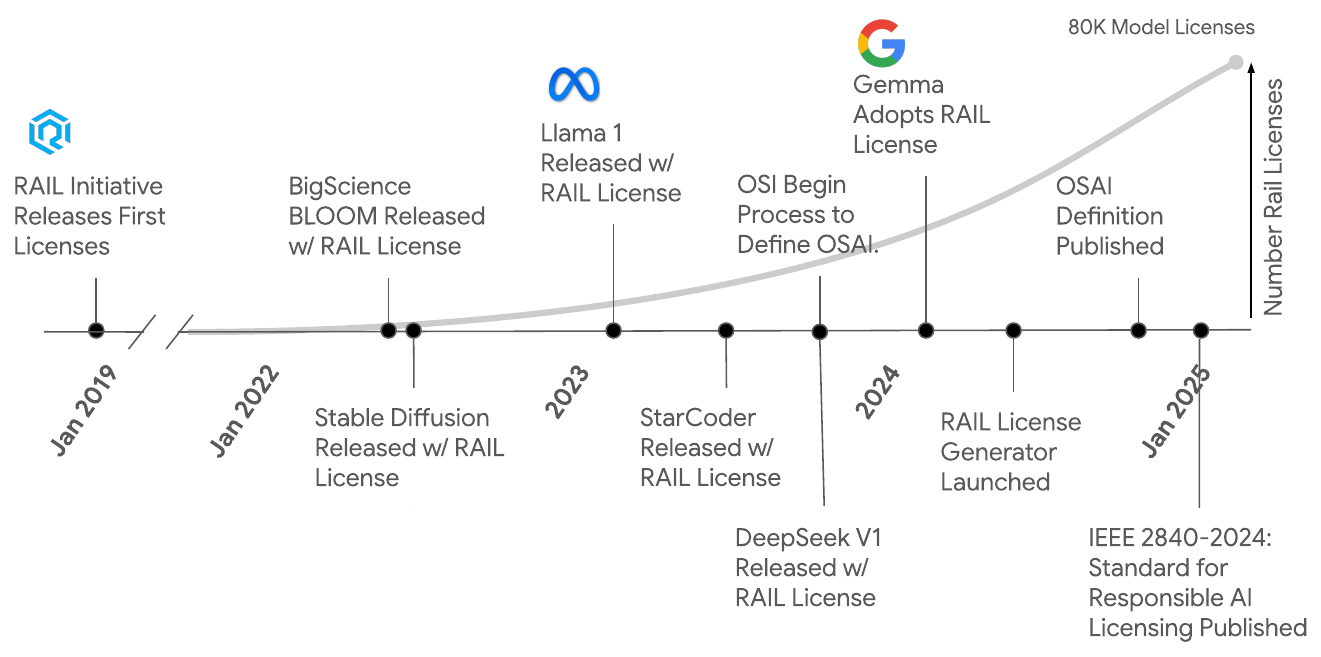}
\caption{
\textbf{RAIL License Release Timeline.}
Notable milestones in the adoption and standardization of RAIL and Open Source AI licenses.}
\label{fig:license_timeline}
\vspace{-0.5cm}
\end{figure}

\begin{figure}
\centering
\includegraphics[width=\textwidth]{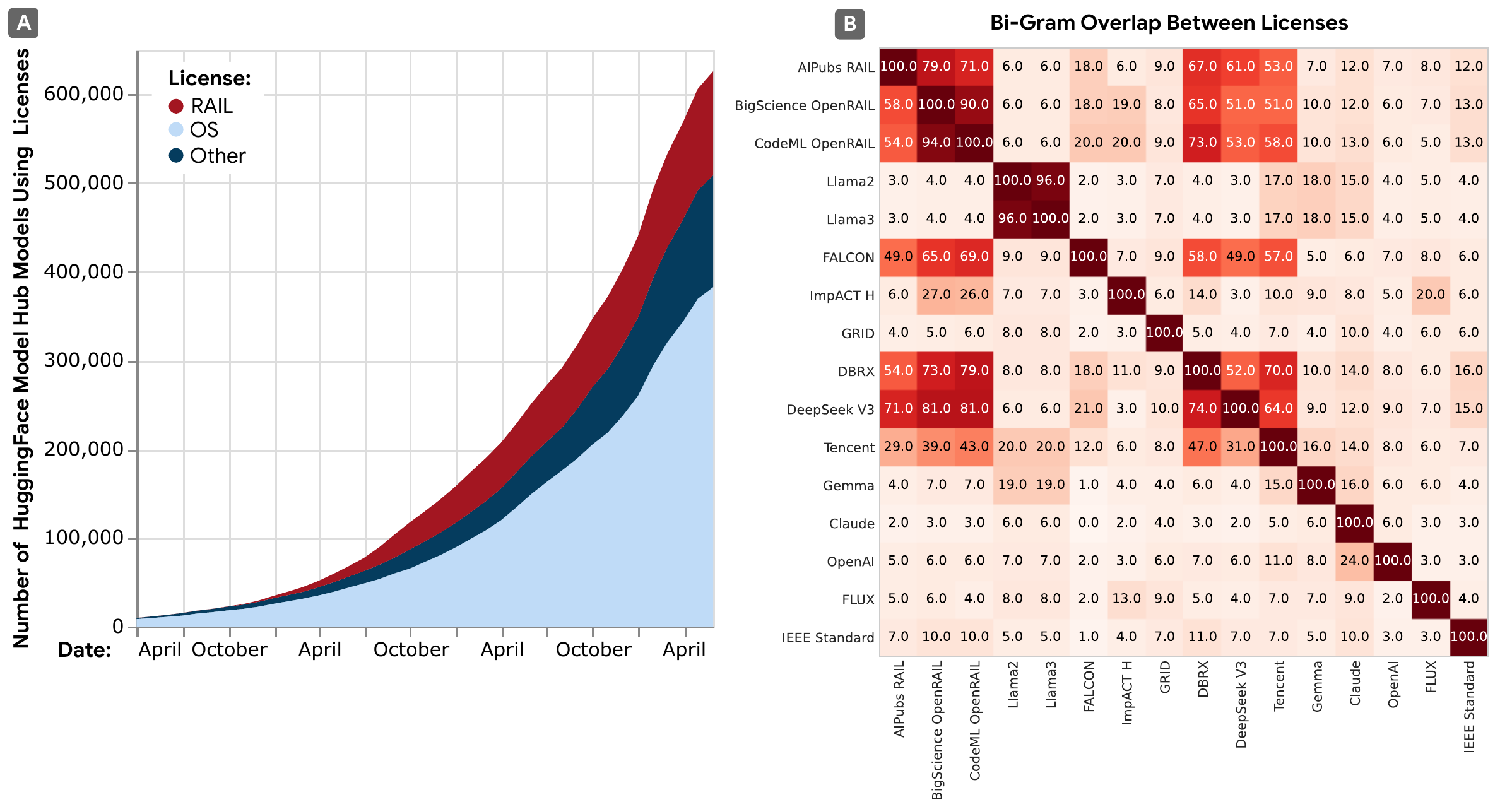}
\caption{
\textbf{License Adoption.} [A] \textbf{Number of Models Licensed.} The number of RAIL, OS and other licensed models on the HuggingFace Model Hub from 2023 to 2025. All other models did not have a license. [B] \textbf{Bi-gram Overlap in Behavioral-use clauses.} The percentage of two-grams that overlap between clauses. There is significant overlap in the text content of certain licenses and most licenses have at least 10\% overlap with at least one other license.
}
\label{fig:license_adoption}
\end{figure}

Next, we performed a large-scale analysis the models hosted on the HuggingFace Hub to identify the licenses associated with AI models. We analyzed 1.7M model repos of these almost 650,000 had licenses. As Figure \ref{fig:license_adoption}[A] shows, RAIL licenses continue to represent a sizable minority of projects compared to Open Source licensed projects (those using the MIT License, Apache License 2.0, Berkeley Software Distribution (BSD) or GNU General Public License (GPL) families of licenses).
RAIL licenses are used for 12.1\% of models, compared to 61.5\% for Open Source.  The timeline in Fig.~\ref{fig:license_timeline} shows some of notable milestone in RAIL adoption and standardization.

\textbf{How consistent are the behavioral-use clauses across these licenses?} A motivation for the RAIL License Generator was the observation that customized licenses were desired~\cite{mcduff2024position}. With the proliferation of behavioral-use licenses created by other parties not only have clauses been added/removed but also clause texts have been changed. Table~\ref{tab:license_clauses} shows the $25$ license clauses in the RAIL License Generator how they appear in some recent RAIL Licenses.
The last column in the table indicates how many (in \%) of the licenses generated  by our license generator include those clauses.

\begin{landscape}
\begin{table*}[ht]
\vspace{-10ex}
	\centering
	\scriptsize
	\setlength\tabcolsep{4pt} 
 \begin{tabular}{ccp{12cm}ccccccccccccccc|c}
	& & \textbf{Behavioral Restrictions} & \rot{\textbf{AIPubs RAIL}} & \rot{\textbf{BigSci. OpenRAIL}} & \rot{\textbf{CodeML OpenRAIL}} & \rot{\textbf{LLaMA 2}} & \rot{\textbf{FALCON}} & \rot{\textbf{ImpACT L/M/H}} & \rot{\textbf{GRID}} & \rot{\textbf{DBRX}} & \rot{\textbf{DeepSeek}} & \rot{\textbf{Tencent}}  
	& \rot{\textbf{Gemma}} & \rot{\textbf{Claude}} & \rot{\textbf{OpenAI}} & \rot{\textbf{FLUX}} & \rot{\textbf{IEEE P2840}} & \rot{\textbf{\% License Gen.}} \\ 
 \toprule 
      \parbox[t]{2mm}{\multirow{5}{*}{\rotatebox[origin=c]{90}{\textbf{Discrim.}}}} & \MandatoryCircle  &
    (1) To discriminate or exploit individuals or groups based on legally protected characteristics and/or vulnerabilities & \cmark & \cmark & \cmark & \cmark & & & \cmark & \cmark & \cmark & \cmark & & \cmark & \cmark & & \cmark & 100\% \\
    & \MandatoryCircle & (2) For purposes of administration of justice, law enforcement, immigration, or asylum processes, such as predicting that a natural person will commit a crime or the likelihood thereof. &  & \cmark & \cmark & & & \cmark & \cmark & \cmark & & & & & \cmark & & \cmark & 100\% \\
    & & (3) To defame, disparage or otherwise harass or exploit others. & \cmark & \cmark & \cmark & \cmark & \cmark & & \cmark & \cmark & \cmark & \cmark & \cmark & \cmark & \cmark & & & 24\% \\
   & & (4) To engage in, promote, incite, or facilitate discrimination or other unlawful or harmful conduct in the provision of employment, employment benefits, credit, housing, or other essential goods and services. &  &  &  &  &  & &  & &  & & & & & & & 18\% \\
\\
   \toprule
   \parbox[t]{2mm}{\multirow{5}{*}{\rotatebox[origin=c]{90}{\textbf{Disinformation}}}}  & \MandatoryCircle & (5) To create, present or disseminate verifiably false or misleading information for economic gain or to intentionally deceive the public, including creating false impersonations of natural persons. & \cmark & \cmark & \cmark & \cmark & & & \cmark & \cmark & \cmark & \cmark & \cmark & \cmark & \cmark & & \cmark & 100\% \\
    & \MandatoryCircle & (6) To synthesize or modify a natural person's appearance, voice, or other individual characteristics, unless prior informed consent of said natural person is obtained & \cmark & \cmark & \cmark & \cmark & & & \cmark & & & & \cmark & \cmark & & & \cmark & 100\%\\
    &  & (7) To generate or disseminate information (including - but not limited to - images, code, posts, articles), and place the information in any public context without expressly and intelligibly disclaiming that the information and/or content is machine generated) &  \cmark & \cmark & \cmark & \cmark & & \cmark & \cmark & \cmark & & \cmark & \cmark & \cmark & \cmark & & & 18\%  \\
       &  \MandatoryCircle & (8) To autonomously interact with a natural person, in text or audio format, unless disclosure and consent is given prior to interaction that the system engaging in the interaction is not a natural person. &   &  & &  & &  &  & & & & & & \cmark & & \cmark & 100\%\\
       
       & & (9) To defame or harm a natural person's reputation, such as by generating, creating, promoting, or spreading defamatory content (statements, images, or other content).  &   &  &  &  & &  &  & & & & \cmark & & \cmark & & & 18\% \\
    \toprule
    \parbox[t]{2mm}{\multirow{2}{*}{\rotatebox[origin=c]{90}{\textbf{Legal}}}} & \MandatoryCircle   & (10) To engage or enable fully automated decision-making that creates, modifies or terminates a binding, enforceable obligation between entities; whether these include natural persons or not. & \cmark & \cmark & \cmark & & & \cmark & \cmark  & \cmark & \cmark & \cmark & \cmark & \cmark & & & \cmark & 100\%\\
    & & (11) In any way that violates any applicable national, federal, state, local or international law or regulation. & \cmark & \cmark & \cmark & \cmark & \cmark & & \cmark & \cmark & \cmark & \cmark & \cmark & \cmark & \cmark & \cmark& & 23\% \\
    
     & \MandatoryCircle & (12) To engage or enable fully automated decision-making that adversely impacts a natural person's legal rights without expressly and intelligibly disclosing the impact to such natural person and providing an appeal process. &  & &  &  &  & &  & & & & \cmark & & \cmark & & \cmark & 100\% \\
    \toprule
    \parbox[t]{2.5mm}{\multirow{3}{*}{\rotatebox[origin=c]{90}{\textbf{Privacy}}}} & \MandatoryCircle  & (13) To utilize personal information to infer additional personal information about a natural person, including but not limited to legally protected characteristics, vulnerabilities or categories; unless informed consent from the data subject to collect said inferred personal information for a stated purpose and defined duration is received. & & & & \cmark & & & & & & & \cmark & \cmark & \cmark & & \cmark & 100\% \\
    & & (14) To generate or disseminate personal identifiable information that can be used to harm an individual or to invade the personal privacy of an individual.  & \cmark  & \cmark  & \cmark & \cmark & & & \cmark  & \cmark & \cmark & \cmark &\cmark & \cmark & \cmark & & & 20\% \\
    &  & (15) To engage in, promote, incite, or facilitate the harassment, abuse, threatening, or bullying of individuals or groups of individuals. & \cmark & \cmark & \cmark & \cmark & \cmark & & \cmark & & & & \cmark & \cmark & \cmark & & & 21\% \\
    \toprule
    \parbox[t]{2mm}{\multirow{3}{*}{\rotatebox[origin=c]{90}{\textbf{Health}}}} & \MandatoryCircle  & (16) To provide medical advice or make clinical decisions without necessary (external) accreditation of the system; unless the use is (i) in an internal research context with independent and accountable oversight and/or (ii) with medical professional oversight that is accompanied by any related compulsory certification and/or safety/quality standard for the implementation of the technology. & & \cmark & \cmark & & \cmark & & &\cmark & & & \cmark & & \cmark & & \cmark & 100\%\\
    &  & (17) In connection with any activities that present a risk of death or bodily harm to individuals, including self-harm or harm to others, or in connection with regulated or controlled substances.  & & & & \cmark & & & & & & \cmark & \cmark &\cmark & \cmark & & & 14\% \\
    & & (18) To provide medical advice and medical results interpretation without external, human validation of such advice or interpretation. & & & & & & & & & & & \cmark & & & & & 14\% \\
    & & (19) In connection with activities that present a risk of death or bodily harm to individuals, including inciting or promoting violence, abuse, or any infliction of bodily harm. & & & & \cmark & & & & & & & \cmark & \cmark & \cmark & & & 13\% \\
    \toprule
    \parbox[t]{2mm}{\multirow{3}{*}{\rotatebox[origin=c]{90}
    {\textbf{Military}}}} & \MandatoryCircle & (20) For weaponry or warfare. & & & & & & &\cmark & & \cmark & \cmark & & \cmark & & \cmark & \cmark & 100\% \\
    & & (21) For purposes of building or optimizing military weapons or in the service of nuclear proliferation or nuclear weapons technology. & & & &\cmark & & \cmark & & & & \cmark & & & & \cmark&  & 17\% \\
    & & (22) For purposes of military surveillance, including any research or development relating to military surveillance. & & & &\cmark & & \cmark & & & & \cmark  & & & & \cmark& & 16\% \\
    \toprule
    \parbox[t]{2mm}{\multirow{2}{*}{\rotatebox[origin=c]{90}{\textbf{Other}}}} & & (23) Generate/disseminate malware/ransomware or other content for the purpose of harming electronic systems. & & & \cmark & & & & & \cmark & & \cmark & \cmark & \cmark& \cmark & & & 19\% \\
    & & (24) To Intentionally deceive or mislead others, including failing to appropriately disclose to end users any known dangers of your system. & & & & \cmark & & & & & & & \cmark & & & & & 25\% \\
    & & (25) In connection with any academic dishonesty, including submitting any informational content or output of a Model as Your own work in any academic setting. & & & & & & & & & & & \cmark& & \cmark & & & 17\% \\
    \bottomrule
  \end{tabular} 
 \caption{\textbf{Summary of Behavioral-Use Clauses.} Clauses included in popular responsible AI licenses.}
	\label{tab:license_clauses}
\end{table*}
\end{landscape}

Notable large language models that have leveraged RAIL licenses include BLOOM~\cite{workshop2022bloom}, StarCoder~\cite{li2023starcoder}, Meta's Llama~\cite{touvron2023llama}, Google's Gemma~\cite{team2024gemma} (acceptable use policy), DeepSeek \cite{liu2024deepseek_v2,liu2024deepseek_v3}, MiniMax-01~\cite{li2025minimax}, Tencent's Hunyuan-large~\cite{sun2024hunyuan}, Falcon ~\cite{almazrouei2023falcon} and Databricks (DBRX).~\footnote{\url{https://www.databricks.com/blog/introducing-dbrx-new-state-art-open-llm}}
Prominent model families with open source licenses without behavioral use clauses include Mistral \cite{jiang2024mixtralexperts}, IBM's Granite \cite{granite2024granite}, Alibaba's Qwen~\cite{bai2023qwen} and Microsoft's Phi \cite{abdin2024phi}.

Figure~\ref{fig:license_adoption}[B] shows the Bi-gram (sequences of two consecutive words) overlap between behavioral-use clauses in different licenses. This analysis shows similarity between almost all the licenses with some containing almost identical language. As an example licenses from the BigScience OpenRAIL-M, Llama and Gemma licenses all contain similar restrictions related to defamation and harassment:

\begin{tcolorbox}
\small
\bigscienceLogo \textbf{BigScience} - \url{https://huggingface.co/spaces/bigscience/license} \\
Clause 1.d. - \textit{"To defame, disparage or otherwise harass others;."}
\end{tcolorbox}

\begin{tcolorbox}
\small
\metaLogo \textbf{Llama 2/3} - \url{https://ai.meta.com/llama/use-policy/} \\
Clause 1.c. - \textit{"Engage in, promote, incite, or facilitate the harassment, abuse, threatening, or bullying of individuals or groups of individual."}
\end{tcolorbox}

\begin{tcolorbox}
\small
\googleLogo \textbf{Gemma} - \url{https://ai.google.dev/gemma/prohibited_use_policy} \\
Clause 2.e.ii. - \textit{"Generation of content that may harm or promote the harm of individuals or a group, such as:
Generating content that promotes or encourages hatred;
Facilitating methods of harassment or bullying to intimidate, abuse, or insult others;"}
\end{tcolorbox}

Apart from the behavioral-use restrictions, licenses such as the Llama license include considerations for commercial-use. Some licenses such as the AIPubs OpenRAIL permit use for Research purposes only. While a study of licenses on dimensions beyond aspects related to behaviorial-use clauses are outside the scope of this work, we encourage the interested readers to review recent work by \cite{10.1145/3589334.3645520, LI2025103330, lemley2025mirage}.

\section{Tools are Need for Tracking Adoption or and Adherence to Licenses}

\textbf{RAIL licenses are being adopted.} The choice to release AI code, models and applications has significant implications for how the technology impacts people and society. Adoption of behavioral use licenses for machine learning models, particularly foundation models, illustrates the desire of researchers to adopt mechanisms to control how these assets are used. Based on analysis of the HuggingFace Model Hub, a sizable portion of models have licenses that include behavioral-use restrictions that reflect how they can be used. The restrictions across these licenses have a large amount of overlap, both in terms of the intention behind clauses and their specific wording. The release of models such as Llama 2~\cite{touvron2023llama}, Gemma~\cite{team2024gemma} and DeepSeek V2 and V3~\cite{liu2024deepseek_v2,liu2024deepseek_v3} accelerated the adoption of certain types of licenses.  Our license generator is a demonstration of the demand for well-tailored licenses: the availability of our tool organically led to the creation of over 300 customized licenses in the space of a year with little additional promotion or advertising.

\textbf{Tools exist for tracking license adoption but could be improved.} As evidenced by the analyses in this work, the task of tracking license adoption has become substantially easier due to GitHub enabling search of a vast number of code repositories, the HuggingFace Model Hub API\footnote{\url{https://huggingface.co/docs/hub/en/models-the-hub}} exposing license categories as metadata and our RAIL License Generator providing a standardized set of language for license clauses. The use of similar language across licenses makes them easier to search. A recently published IEEE standard\footnote{\url{https://standards.ieee.org/ieee/2840/7673/}} could help to further increase the coherence of these licenses making it easier to track adoption. This tooling builds on prior work to build more standardized and transparency licenses for machine learning models \cite{duan2024theyvestolengpllicensedmodel}. The lack of a standardized filetype for licenses (some are txt or markdown files, while others are word documents) limits systematic analysis, especially as many license files are titled ``LICENSE,'' and Hugging Face's Model Hub API does not enable search by license filetype; our custom AI license generator allows the creation of licenses as markdown files. Tools for analyzing license content also exist \cite{10.1145/3589334.3645520}, such as automated text analysis tools and platforms for carrying out qualitative coding and simplifying subsequent analysis \cite{drazewski2021corpus, guarino2021machine}, though increasingly capable language models may reduce the need for such tooling over time.

\textbf{Tools for tracking license adherence are severely lacking.} Identifying violations to license terms is a serious challenge. Existing approaches to Open Source code monitoring broadly fall into three buckets: 1) automated code scanning, 2) manual code reviews and 3) community monitoring and reporting. While existing automated code scanning tools would still be helpful to identify when code that is under a RAIL license is being used, they many be less effective at detecting \emph{how} that code is being used and whether a behavioral restriction is being violated.
The solution to tracking adherence to license clauses is more complex, and will likely need socio-technical solutions.  Taking inspiration from the Open Source Initiative, greater adoption of tools for monitoring public complaints such as those on the Open Source Stack Exchange\footnote{\url{https://opensource.stackexchange.com/}} would provide a way for community reported violations to be escalated.

While larger organization might have the resources to track model use, practitioners from academic or smaller organizations (as is the case form many people who used the license generator) may struggle to identify how their assets are being used and whether any of the license terms are being violated. Novel methods for model provenance, such as model fingerprinting \cite{xu2024instructionalfingerprintinglargelanguage}, or output watermarking, such as robust distortion-free watermarks \cite{kuditipudi2024robustdistortionfreewatermarkslanguage, fernandez2023stablesignaturerootingwatermarks}, may provide a useful substrate for tracking violations of license terms in time. However, the lack of viable methods for tracing the provenance of models or their outputs, coupled with little infrastructure for identifying violations of license terms even when a model and its outputs can be correctly tracked, may lead developers to seek out ways to ``short-circuit'' model performance in certain domains in an effort to promote adherence to behavioral use clauses\cite{zou2024improvingalignmentrobustnesscircuit}.


\section{Alternative Views}

There are alternative views to the position taken in this paper about how to release and license AI technology. For instance, the Open Source Initiative (OSI) launched a co-design process to define Open Source AI definition, partly in response to the community-led the adoption of licenses with behavioural-use clauses.\footnote{\url{https://opensource.org/blog/towards-a-definition-of-open-artificial-intelligence-first-meeting-recap}}$^,$\footnote{ \url{https://opensource.org/ai/open-weights}} The open-source community values the four `freedoms' of free software:\footnote{\url{https://www.gnu.org/philosophy/free-sw.en.html\#four-freedoms}} (i) freedom to run the software for any purpose, (ii) freedom to study how a software works, (iii) freedom to re-distribute, (iv) freedom to distribute modified versions to others. It is therefore obvious how licenses such as those that restrict commercial-use, or impose behavioral-use requirements contradict these freedoms. 

The recently released Open Source AI Definition (OSAID) v1.0\footnote{\url{https://opensource.org/ai}} reinforces the principles of free-software and also makes the distinction between `Open Source AI' and `Open-Weights'. Open Source AI systems need to release training data when possible, release weights of trained models, along with the source code to train and evaluate AI Models. These artifacts are deemed necessary to comply with the four freedoms though the definition makes room for practical considerations such as acknowledging that training data may not always be possible to release (e.g., data including PII). On the other hand, open-weight models refer to models that do not impose restrictions of any kind but they may not be accompanied by the release of code or data. Most models released under popular licenses such as Apache 2.0 and MIT License are thus, Open-Weight Models (including the ones previously refereed to as ``Open-Source'' models in Section \ref{sec:commonlyadoptedailicenses}). In fact, very few models conform to the definition of prescribed by OSAID, for examples models such as OLMo 2\footnote{\url{https://allenai.org/blog/olmo2}} and Google T5 \cite{t5} would not.

\subsection{Open Weights versus RAIL} \label{sec:openweightvsrail}
The tension between unrestricted access and responsible-use of AI is most notable when there have been instances of model providers switching between RAIL and Open Weight releases. For instance, the Stable Diffusion Model first released under a license with behavioral-use restrictions\footnote{\url{https://raw.githubusercontent.com/CompVis/stable-diffusion/main/LICENSE}} and then switched to an open source license, before finally reverting to a RAIL-license. DeepSeek's early models including DeepSeek V3\footnote{\url{https://github.com/deepseek-ai/DeepSeek-V3/blob/main/LICENSE-MODEL}} are licensed with behavioural-use clauses, but its recent DeepSeek R1 model is licensed under an MIT license.\footnote{\url{https://huggingface.co/deepseek-ai/DeepSeek-R1/blob/main/LICENSE}} Model families such as Phi release weights under open source licenses but include guidance to comply with laws and request users to carefully consider the use of the model particularly in high-risk use-cases.\footnote{\url{https://huggingface.co/microsoft/Phi-3.5-vision-instruct}}$^,$\footnote{\url{https://huggingface.co/microsoft/phi-4\#intended-use}} This suggests that model providers are actively thinking about the use of their models but grapple with the uncertainty that the inclusion of responsible-use clauses can bring especially when these have not been standardized or tested in court. 

Considerations on responsible-use have also been applied to artifacts such as datasets -- for instance, the Dolma dataset was initially released under an ImpACT license (a license with behavioral-use restrictions) and then switched to an ODC By 2.0 License,\footnote{\url{https://allenai.org/blog/making-a-switch-dolma-moves-to-odc-by-8f0e73852f44}} which enabled the OLMo family of models (a model trained on this dataset) to be released under an Apache 2.0 license. 





\section{Conclusion}

Increasingly, AI software assets (models, source code, applications) are being released with licenses that include behavioral-use clauses and acceptable-use policies. While the specific combinations of clauses varies, we find that most continue use clauses from a relatively short list (N=25). A field-study of a license generation tool found organic adoption (N=308 licenses created in 12-month) analysis of these licenses found that 50\% of practitioners chose to customize the choice of license clauses, including behavioral-use clause more when licensing models than source code or applications. With the trend of adoption of licenses clearly continuing and the validation of tools to help people create customized licenses, attention needs to move to support for tracking usage of assets and adherence to these licenses.

\bibliography{neurips_refs}

\begin{thebibliography}{10}

\bibitem{sun2024hunyuan}
Hunyuan-large: An open-source moe model with 52 billion activated parameters by tencent.
\newblock {\em arXiv preprint arXiv:2411.02265}, 2024.

\bibitem{abdin2024phi}
Marah Abdin, Jyoti Aneja, Hany Awadalla, Ahmed Awadallah, Ammar~Ahmad Awan, Nguyen Bach, Amit Bahree, Arash Bakhtiari, Jianmin Bao, Harkirat Behl, et~al.
\newblock Phi-3 technical report: A highly capable language model locally on your phone.
\newblock {\em arXiv preprint arXiv:2404.14219}, 2024.

\bibitem{almazrouei2023falcon}
Ebtesam Almazrouei, Hamza Alobeidli, Abdulaziz Alshamsi, Alessandro Cappelli, Ruxandra Cojocaru, Merouane Debbah, Etienne Goffinet, Daniel Heslow, Julien Launay, Quentin Malartic, et~al.
\newblock {Falcon-40B: an open large language model with state-of-the-art performance}.
\newblock {\em Findings of the Association for Computational Linguistics: ACL}, 2023:10755--10773, 2023.

\bibitem{bai2023qwen}
Jinze Bai, Shuai Bai, Yunfei Chu, Zeyu Cui, Kai Dang, Xiaodong Deng, Yang Fan, Wenbin Ge, Yu~Han, Fei Huang, et~al.
\newblock Qwen technical report.
\newblock {\em arXiv preprint arXiv:2309.16609}, 2023.

\bibitem{contractor2022behavioral}
Danish Contractor, Daniel McDuff, Julia~Katherine Haines, Jenny Lee, Christopher Hines, Brent Hecht, Nicholas Vincent, and Hanlin Li.
\newblock Behavioral use licensing for responsible ai.
\newblock In {\em Proceedings of the 2022 ACM Conference on Fairness, Accountability, and Transparency}, pages 778--788, 2022.

\bibitem{noauthor_orcasoundorca-eye-aye_2024}
Ze~Cui.
\newblock orcasound/orca-eye-aye.
\newblock original-date: 2023-03-17T19:11:42Z.

\bibitem{drazewski2021corpus}
Kasper Drazewski, Andrea Galassi, Agnieszka JAB{\L}ONOWSKA, Francesca Lagioia, Marco Lippi, Hans-Wolfgang Micklitz, Giovanni Sartor, Giacomo Tagiuri, and Paolo Torroni.
\newblock A corpus for multilingual analysis of online terms of service.
\newblock Association for Computational Linguistics, 2021.

\bibitem{10.1145/3589334.3645520}
Moming Duan, Qinbin Li, and Bingsheng He.
\newblock Modelgo: A practical tool for machine learning license analysis.
\newblock In {\em Proceedings of the ACM Web Conference 2024}, WWW '24, page 1158–1169, New York, NY, USA, 2024. Association for Computing Machinery.

\bibitem{duan2024theyvestolengpllicensedmodel}
Moming Duan, Rui Zhao, Linshan Jiang, Nigel Shadbolt, and Bingsheng He.
\newblock "they've stolen my gpl-licensed model!": Toward standardized and transparent model licensing, 2024.

\bibitem{fernandez2023stablesignaturerootingwatermarks}
Pierre Fernandez, Guillaume Couairon, Hervé Jégou, Matthijs Douze, and Teddy Furon.
\newblock The stable signature: Rooting watermarks in latent diffusion models, 2023.

\bibitem{florent_bartoccioni_vavim_2025}
{Florent Bartoccioni}, {Elias Ramzi}, {Victor Besnier}, {Shashanka Venkataramanan}, {Tuan-Hung Vu}, {Yihong Xu}, {Loick Chambon}, {Spyros Gidaris}, {Serkan Odabas}, {David Hurych}, {Renaud Marlet}, {Alexandre Boulch}, {Mickael Chen}, {Eloi Zablocki}, {Andrei Bursuc}, {Eduardo Valle}, and {Matthieu Cord}.
\newblock {VaViM} and {VaVAM}: Autonomous driving through video generative modeling, 2024.
\newblock original-date: 2024-02-09T07:46:03Z.

\bibitem{granite2024granite}
IBM Granite~Team.
\newblock Granite 3.0 language models, 2024.

\bibitem{guarino2021machine}
Alfonso Guarino, Nicola Lettieri, Delfina Malandrino, and Rocco Zaccagnino.
\newblock A machine learning-based approach to identify unlawful practices in online terms of service: analysis, implementation and evaluation.
\newblock {\em Neural Computing and Applications}, 33:17569--17587, 2021.

\bibitem{isobe_amd-hummingbird_2025}
Takashi Isobe, He~Cui, Dong Zhou, Mengmeng Ge, Dong Li, and Emad Barsoum.
\newblock {AMD}-hummingbird: Towards an efficient text-to-video model, 2025.

\bibitem{jiang2024mixtralexperts}
Albert~Q. Jiang, Alexandre Sablayrolles, Antoine Roux, Arthur Mensch, Blanche Savary, Chris Bamford, Devendra~Singh Chaplot, Diego de~las Casas, Emma~Bou Hanna, Florian Bressand, Gianna Lengyel, Guillaume Bour, Guillaume Lample, Lélio~Renard Lavaud, Lucile Saulnier, Marie-Anne Lachaux, Pierre Stock, Sandeep Subramanian, Sophia Yang, Szymon Antoniak, Teven~Le Scao, Théophile Gervet, Thibaut Lavril, Thomas Wang, Timothée Lacroix, and William~El Sayed.
\newblock Mixtral of experts, 2024.

\bibitem{joshi2024factorizephys}
Jitesh Joshi, Sos Agaian, and Youngjun Cho.
\newblock Factorizephys: Matrix factorization for multidimensional attention in remote physiological sensing.
\newblock In {\em The Thirty-eighth Annual Conference on Neural Information Processing Systems}, 2024.

\bibitem{kuditipudi2024robustdistortionfreewatermarkslanguage}
Rohith Kuditipudi, John Thickstun, Tatsunori Hashimoto, and Percy Liang.
\newblock Robust distortion-free watermarks for language models, 2024.

\bibitem{lemley2025mirage}
Mark~A. Lemley and Peter Henderson.
\newblock The mirage of artificial intelligence terms of use restrictions.
\newblock Research Paper 2025-04, Princeton University Program in Law \& Public Affairs, 2024.

\bibitem{li2025minimax}
Aonian Li, Bangwei Gong, Bo~Yang, Boji Shan, Chang Liu, Cheng Zhu, Chunhao Zhang, Congchao Guo, Da~Chen, Dong Li, et~al.
\newblock Minimax-01: Scaling foundation models with lightning attention.
\newblock {\em arXiv preprint arXiv:2501.08313}, 2025.

\bibitem{LI2025103330}
Peihao Li, Jie Huang, and Shuaishuai Zhang.
\newblock Licensenet: Proactively safeguarding intellectual property of ai models through model license.
\newblock {\em Journal of Systems Architecture}, 159:103330, 2025.

\bibitem{li2023starcoder}
Raymond Li, Loubna~Ben Allal, Yangtian Zi, Niklas Muennighoff, Denis Kocetkov, Chenghao Mou, Marc Marone, Christopher Akiki, Jia Li, Jenny Chim, et~al.
\newblock Starcoder: may the source be with you!
\newblock {\em arXiv preprint arXiv:2305.06161}, 2023.

\bibitem{liu2024deepseek_v2}
Aixin Liu, Bei Feng, Bin Wang, Bingxuan Wang, Bo~Liu, Chenggang Zhao, Chengqi Dengr, Chong Ruan, Damai Dai, Daya Guo, et~al.
\newblock Deepseek-v2: A strong, economical, and efficient mixture-of-experts language model.
\newblock {\em arXiv preprint arXiv:2405.04434}, 2024.

\bibitem{liu2024deepseek_v3}
Aixin Liu, Bei Feng, Bing Xue, Bingxuan Wang, Bochao Wu, Chengda Lu, Chenggang Zhao, Chengqi Deng, Chenyu Zhang, Chong Ruan, et~al.
\newblock Deepseek-v3 technical report.
\newblock {\em arXiv preprint arXiv:2412.19437}, 2024.

\bibitem{jiang_liu_instella_2025}
Jiang Liu, Jialian Wu, Xiaodong Yu, Sudhanshu~Ranjan Prakamya~Mishra, Zicheng Liu, Chaitanya Manem, Yusheng Su, Pratik~Prabhanjan Brahma, Gowtham Ramesh, Ximeng Sun, Ze~Wang, and Emad Barsoum.
\newblock Instella: Fully open language models with stellar performance, 2025.

\bibitem{longpreresponsible}
Shayne Longpre, Stella Biderman, Alon Albalak, Hailey Schoelkopf, Daniel McDuff, Sayash Kapoor, Kevin Klyman, Kyle Lo, Gabriel Ilharco, Nay San, et~al.
\newblock The responsible foundation model development cheatsheet: A review of tools \& resources.
\newblock {\em Transactions on Machine Learning Research}.

\bibitem{mcduff2024position}
Daniel McDuff, Tim Korjakow, Scott Cambo, Jesse~Josua Benjamin, Jenny Lee, Yacine Jernite, Carlos~Mu{\~n}oz Ferrandis, Aaron Gokaslan, Alek Tarkowski, Joseph Lindley, et~al.
\newblock Position: Standardization of behavioral use clauses is necessary for the adoption of responsible licensing of ai.
\newblock In {\em International Conference on Machine Learning}, pages 35255--35266. PMLR, 2024.

\bibitem{t5}
Colin Raffel, Noam Shazeer, Adam Roberts, Katherine Lee, Sharan Narang, Michael Matena, Yanqi Zhou, Wei Li, and Peter~J. Liu.
\newblock Exploring the limits of transfer learning with a unified text-to-text transformer.
\newblock {\em J. Mach. Learn. Res.}, 21(1), January 2020.

\bibitem{rombach2022high}
Robin Rombach, Andreas Blattmann, Dominik Lorenz, Patrick Esser, and Bj{\"o}rn Ommer.
\newblock High-resolution image synthesis with latent diffusion models.
\newblock In {\em Proceedings of the IEEE/CVF conference on computer vision and pattern recognition}, pages 10684--10695, 2022.

\bibitem{noauthor_salmon-computer-visionsalmon-computer-vision_2025}
SalmonVision.
\newblock Salmon-computer-vision/salmon-computer-vision.
\newblock original-date: 2020-09-14T21:14:11Z.

\bibitem{ximeng_sun_instella-vl-1b_2025}
Ximeng Sun, Aditya Singh, Gowtham Ramesh, Jiang Liu, Ze~Wang, Sudhanshu Ranjan, Pratik~Prabhanjan Brahma, Prakamya Mishra, Jialian Wu, Xiaodong Yu, Yusheng Su, Emad Barsoum, and Zicheng Liu.
\newblock Instella-{VL}-1b: First {AMD} vision language model, 2025.

\bibitem{foundationaiteam2025cuberobloxview3d}
Foundation~AI Team, Kiran Bhat, Nishchaie Khanna, Karun Channa, Tinghui Zhou, Yiheng Zhu, Xiaoxia Sun, Charles Shang, Anirudh Sudarshan, Maurice Chu, Daiqing Li, Kangle Deng, Jean-Philippe Fauconnier, Tijmen Verhulsdonck, Maneesh Agrawala, Kayvon Fatahalian, Alexander Weiss, Christian Reiser, Ravi~Kiran Chirravuri, Ravali Kandur, Alejandro Pelaez, Akash Garg, Michael Palleschi, Jessica Wang, Skylar Litz, Leon Liu, Anying Li, David Harmon, Derek Liu, Liangjun Feng, Denis Goupil, Lukas Kuczynski, Jihyun Yoon, Naveen Marri, Peiye Zhuang, Yinan Zhang, Brian Yin, Haomiao Jiang, Marcel van Workum, Thomas Lane, Bryce Erickson, Salil Pathare, Kyle Price, Anupam Singh, and David Baszucki.
\newblock Cube: A roblox view of 3d intelligence, 2025.

\bibitem{team2024gemma}
Gemma Team, Thomas Mesnard, Cassidy Hardin, Robert Dadashi, Surya Bhupatiraju, Shreya Pathak, Laurent Sifre, Morgane Rivi{\`e}re, Mihir~Sanjay Kale, Juliette Love, et~al.
\newblock Gemma: Open models based on gemini research and technology.
\newblock {\em arXiv preprint arXiv:2403.08295}, 2024.

\bibitem{team2024gemma2}
Gemma Team, Morgane Riviere, Shreya Pathak, Pier~Giuseppe Sessa, Cassidy Hardin, Surya Bhupatiraju, L{\'e}onard Hussenot, Thomas Mesnard, Bobak Shahriari, Alexandre Ram{\'e}, et~al.
\newblock Gemma 2: Improving open language models at a practical size.
\newblock {\em arXiv preprint arXiv:2408.00118}, 2024.

\bibitem{touvron2023llama}
Hugo Touvron, Louis Martin, Kevin Stone, Peter Albert, Amjad Almahairi, Yasmine Babaei, Nikolay Bashlykov, Soumya Batra, Prajjwal Bhargava, Shruti Bhosale, et~al.
\newblock Llama 2: Open foundation and fine-tuned chat models.
\newblock {\em arXiv preprint arXiv:2307.09288}, 2023.

\bibitem{vandal_global_2024}
Thomas~J Vandal, Kate Duffy, Daniel {McDuff}, Yoni Nachmany, and Chris Hartshorn.
\newblock Global atmospheric data assimilation with multi-modal masked autoencoders.
\newblock 2024.

\bibitem{workshop2022bloom}
BigScience Workshop, Teven~Le Scao, Angela Fan, Christopher Akiki, Ellie Pavlick, Suzana Ili{\'c}, Daniel Hesslow, Roman Castagn{\'e}, Alexandra~Sasha Luccioni, Fran{\c{c}}ois Yvon, et~al.
\newblock Bloom: A 176b-parameter open-access multilingual language model.
\newblock {\em arXiv preprint arXiv:2211.05100}, 2022.

\bibitem{xu2024instructionalfingerprintinglargelanguage}
Jiashu Xu, Fei Wang, Mingyu~Derek Ma, Pang~Wei Koh, Chaowei Xiao, and Muhao Chen.
\newblock Instructional fingerprinting of large language models, 2024.

\bibitem{zou2024improvingalignmentrobustnesscircuit}
Andy Zou, Long Phan, Justin Wang, Derek Duenas, Maxwell Lin, Maksym Andriushchenko, Rowan Wang, Zico Kolter, Matt Fredrikson, and Dan Hendrycks.
\newblock Improving alignment and robustness with circuit breakers, 2024.

\end{thebibliography}
\bibliographystyle{plain}

\appendix
\section{Broader Impacts}

Licenses have broad implications. The open source community has contributed substantially to the development of software and to the progress in artificial intelligence. There are reasons that restrictive licenses might impact research and development and justified fears that they could slow down progress or cause confusion about what is and what is not acceptable with a given asset. We are not arguing behavioral use clauses to be a required component of licenses, but rather that they are an option available to practitioners. In this paper term OpenRAIL and RAIL are used specifically to distinguish these license types from Open Source licenses. Our analyses suggest there has formed a community-driven consensus around the types of behavioral-use clauses are needed.

\end{document}